\documentstyle[aps,prl,epsfig,multicol]{revtex}

\def\CC{{\rm\kern.24em \vrule width.04em height1.46ex depth-.07ex
\kern-.30em C}}

\begin{document}
\draft
\title{Coherence-Preserving Quantum Bits}
\author{Dave Bacon$^{1,2}$, Kenneth R. Brown$^{1}$, and K. Birgitta Whaley$^1$}
\address{Departments of Chemistry$^1$ and Physics$^2$, University of
California, Berkeley 94704}
\date{\today}
\maketitle

\begin{abstract}
Real quantum systems couple to their environment and lose their intrinsic
quantum nature through the process known as decoherence.  Here we present a
method for minimizing decoherence by making it energetically
unfavorable.  We present a Hamiltonian made up solely of two-body interactions
between four two-level systems (qubits) which has a two-fold degenerate ground
state.  This degenerate ground state has the property that any decoherence
process acting on an individual physical qubit {\em must supply energy from
the bath to the system}.
Quantum information can be encoded into the degeneracy of the ground state and
such coherence-preserving qubits will then be robust to local decoherence at low
bath temperatures.  We show how this quantum information can be universally manipulated and indicate how this approach may be applied to a quantum dot quantum computer.
\end{abstract}

\pacs{PACS Numbers: 03.67.Lx,03.65.Yz, 03.67.-a}

\begin{multicols}{2}

\narrowtext

One of the most severe experimental difficulties in quantum information
processing is the fragile nature of quantum information.  Every real quantum
system is an open system which readily couples to its environment. This
coupling causes the quantum information in the system to become entangled with
its environment, which in turn results in the system information losing its
intrinsic quantum nature. This process is known as decoherence. Circumvention
of this {\em decoherence problem} has been shown to be theoretically possible
with the development of the theory of fault-tolerant quantum error correction
\cite{Preskill:97a}.  The set of requirements to reach the threshold for such
fault-tolerant quantum computation is, however, extremely daunting.  In this
Letter we present a quantum informatic method for suppressing the detrimental
effects of decoherence, while at the same time allowing for robust manipulation
of the quantum information, in the hope that this method will aid in breeching
the threshold for robust quantum computation \cite{FTpeople}.

In the absence of coupling between a system and its environment, the system and
environment have separate temporal evolutions determined by their individual
energy spectra.  When a small interaction (relative to these energy scales) is
switched on between the two, the resulting evolution is dominated by pathways
that conserve the energy of the unperturbed system plus environment (rotating
wave approximation, see \cite{scully}).  Under the assumption of such a
perturbative interaction, energetics play a key role in determining the rate of
decoherence processes.  Such energy conserving decoherence has three possible
forms: energy is supplied from the system to the environment (cooling), energy
is supplied from the environment to the system (heating), or no energy is
exchanged at all (non-dissipative).  Thus, even when the environment is a heat
bath at zero temperature, cooling and especially non-dissipative interactions
can be a major source of decoherence.

The spirit of our approach to reducing decoherence is to force all {\em
reasonable} decoherence mechanisms to be interactions which heat the system,
such that at low bath temperatures decoherence is energetically
suppressed. This is done by encoding into logical qubits which are the ground state of a particular engineered Hamiltonian.  While all dissipative and dephasing processes act on the physical qubits, the only source of decoherence on the encoded qubits derives from
non-energy conserving decoherence pathways, which are by definition perturbatively weak. In particular, we will show the existence of a degenerate
collective ground state of pairwise interacting two-level systems (qubits),
which possesses the property that any local operation on an individual physical
qubit must take the system out of this collective ground state. Quantum information can
be encoded into the degeneracy of this ground state, to make an encoded qubit that is protected from any
local decoherence which cannot overcome the established energy gap.

\begin{figure}[h]

\epsfig{figure=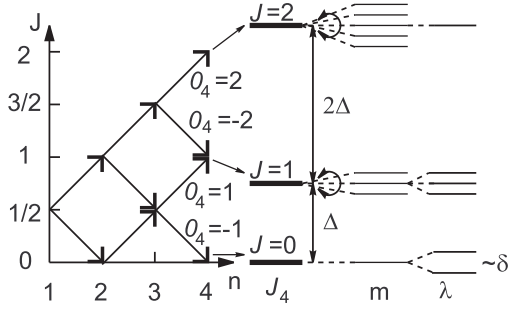,width=220pt} \caption{Diagram showing formation of the $|\lambda,J_n,m\rangle$ states.  The degeneracy index $\lambda$ of a given $J$ irreducible representation can be found by counting the number of paths
which start with a spin-$1/2$ particle and which build up a total spin of $J$
using standard addition of angular momenta.  Thus, each path in
this figure starting from $n=1$, $J_1=1/2$ is in one-to-one correspondence with
a degeneracy index $\lambda$ of a given $J$ irrep. $O_n$ is the eigenvalue of
${\bf O}_n$ for the final step of this pathway. The allowed $\Delta J$ transitions are shown as double-ended
arrows between the energy levels of ${\bf H}_0^{(4)}$ (see text for
definition).  Shown on the right are the $\lambda$ and $m$ degeneracies of the $J_4$ levels.  The energy difference $\delta$ corresponding to a computation
on the supercoherent qubit will split the $\lambda$ degeneracy. } \label{fig1}
\end{figure}

{\em Collective spin operations.}---Let ${\cal H}_n=(\CC^2)^{\otimes n}$ be a
Hilbert space of $n$ qubits, and let ${\bf s}_\alpha^{(i)}$ be the $\alpha$th
Pauli spin operator acting on the $i$th qubit tensored with identity on all
other qubits.  The ${\bf s}_\alpha^{(i)}$ satisfy $[{\bf s}_\alpha^{(j)},{\bf s}_\beta^{(k)}]=i\delta_{jk}
\epsilon_{\alpha \beta \gamma} {\bf s}_\gamma^{(j)}$ and $ \{ {\bf
s}_\alpha^{(j)}, {\bf s}_\beta^{(k)}  \} = {1 \over 2} \delta_{jk}
\delta_{\alpha \beta} {\bf I} + 2(1-\delta_{jk}){\bf s}_\alpha^{(j)} {\bf
s}_\beta^{(k)}$.  We define the $k$th partial collective spin operators on the
$n$ qubits, ${\bf S}_\alpha^{(k)} = \sum_{i=1}^k {\bf s}_\alpha^{(i)}$. The
total collective spin operators acting on all $n$ qubits, ${\bf
S}_{\alpha}^{(n)}$, form a Lie algebra ${\cal L}$ which provides a
representation of
the Lie algebra $su(2)$: $[{\bf S}_{\alpha}^{(n)},{\bf S}_\beta^{(n)}]=i
\epsilon_{\alpha \beta \gamma} {\bf S}_{\gamma}^{(n)}$.  Thus ${\cal L}$ can be
decomposed in a direct product of irreducible representations (irreps) of
$su(2)$, ${\cal L} \simeq \bigoplus_{J=0,1/2}^{n/2} \bigoplus_{k=1}^{n_J} {\cal
L}_{2J+1}$, where ${\cal L}_{2J+1}$ is the $2J+1$ dimensional irrep of $su(2)$
which appears with a multiplicity $n_J$.  If we let $({\bf J}_d)_\alpha$ be the
operators of the $d$ dimensional irrep of $su(2)$, then there exists a basis for the
total collective spin operators such that ${\bf S}_{\alpha}^{(n)}=
\bigoplus_{J=0,1/2}^{n/2} {\bf I}_{n_J} \otimes ({\bf J}_{2J+1})_{\alpha}$.
Corresponding to this decomposition of ${\bf S}_{\alpha}^{(n)}$, the
Hilbert space ${\cal H}$ can be decomposed into states $|\lambda,J_n,m\rangle$
classified by quantum numbers labeling the irrep, $J_n$,
the degeneracy index of the irrep, $\lambda$,
and an additional internal degree of freedom, $m$.
A complete set of
commuting operators consistent with this decomposition and providing explicit
values for these
labels is given by $B_\alpha= \{ (\vec{\bf S}^{(1)})^2, (\vec{\bf
S}^{(2)})^2,\dots, (\vec{\bf S}^{(n-1)})^2, (\vec{\bf S}^{(n)})^2, {\bf
S}^{(n)}_\alpha \}$ \cite{Kempe:00}.  Therefore a basis for the entire Hilbert
space is given by $|J_1,J_2,\dots,J_{n-1},J_n,m_\alpha \rangle$, with
$(\vec{\bf S}^{(k)})^2|J_1,\dots,J_n,m_\alpha\rangle = J_k (J_k+1)
|J_1,\dots,J_n,m_\alpha\rangle$ and ${\bf
S}_\alpha^{(n)}|J_1,\dots,J_n,m_\alpha\rangle = m_\alpha
|J_1,\dots,J_n,m_\alpha\rangle$.  The degeneracy index $\lambda$ of a
particular
irrep having total collective spin $J_n$ is completely specified by
the set of partial collective spin eigenvalues $J_k$, $k<n$:
$\lambda \equiv ( J_1,\dots, J_{n-1} )$.
This degeneracy is simply due to the ($n_J$) different possible ways of
constructing a spin-$J_n$ out of $n$ qubits.  In Fig.~\ref{fig1} we present a
graphical method for understanding this degeneracy of the irreps.  The
internal quantum number $m_{\alpha}$ is the total spin projection
along axis $\alpha$.

The $|\lambda,J_n,m\rangle$ states have a particular clean property for
decoherence mechanisms which couple collectively to the system.  Quantum
information encoded into the degeneracy $|\lambda\rangle$ of these states is
immune to collective decoherence. This information inhabits a
decoherence-free (noiseless) subsystem \cite{Zanardi,Zanardi:97,Knill:00,Kempe:00}.
Non-collective or local errors can still adversely affect decoherence-free
subsystems\cite{LidarBacon}.  In this paper we consider the action of
independent errors acting on a code derived from decoherence free states and we show that these errors can be
suppressed by suitable construction of the energy spectum.  The decoherence-free property of the encoded states is
retained in our approach. However, the method we present here deals with
independent errors: as such, it can be used to reduce these errors
irrespective of the existence of collective decoherence.

 {\em Collective Hamiltonian.}---The Hamiltonian ${\bf H}_0^{(n)}={\Delta \over 2}(\vec{\bf S}^{(n)})^2$ has
eigenvalues ${\Delta \over 2} J_n (J_n+1)$, with corresponding eigenstates
$|\lambda,J_n,m_\alpha \rangle$. Thus the (possibly degenerate) ground state of
such a Hamiltonian is given by the lowest $J_n$ states for a particular $n$.
For $n$ even, these states have $J_n=0$, and for $n$ odd they have $J_n=1/2$.
Furthermore, ${\bf H}_0^{(n)}$ can be constructed from two-qubit interactions
%BW 05/03
alone: ${\bf H}_0^{(n)} = {\Delta \over 2}  \left(\sum_{i \neq j=1}^n \vec{\bf
s}^{(i)} \cdot \vec{\bf s}^{(j)} + {3 n \over 4} {\bf I} \right)$.   Thus we
see that ${\bf H}_0^{(n)}$ is nothing more than the Heisenberg coupling
$\vec{\bf s}^{(i)} \cdot \vec{\bf s}^{(j)}$ acting with equal magnitude between
every pair of qubits  (${\bf I}$ is an irrelevant energy shift).

%The identity component of ${\bf H}_0^{(n)}$ produces only a trivial
%global phase on the system and is not relevant to our discussion.

{\em Effect of single qubit operators.}---${\bf H}_0^{(n)}$ has a highly
degenerate spectrum, with energies determined by $J_n$.  To determine the
effect of single qubit operations on these states first consider the effect
of a single qubit operation on the $n$th qubit, ${\bf s}_\alpha^{(n)}$.  Since
$[{\bf s}_\alpha^{(n)},(\vec{\bf S}^{(k)})^2]=0$ for $k<n$, we see that ${\bf
s}_\alpha^{(n)}$ can not change the degeneracy index $\lambda$ of a state
$|\lambda,J_n,m_\alpha\rangle$.  Let ${\bf O}_n=-{1 \over 4} {\bf I}+(\vec{\bf
S}^{(n)})^2-(\vec{\bf S}^{(n-1)})^2$ (defined for $n>1$).  ${\bf O}_n$
determines which final step is taken in  the addition from qubit $n-1$ to
qubit $n$ (Fig.~\ref{fig1}). If the final step from $J_{n-1}$ to $J_n$ was
taken by adding $1/2$, then the eigenvalue of ${\bf O}_n$ will be
$O_n=J_{n-1}+{1 \over 2}$, while if it was taken by subtracting $1/2$, then
$O_n=-(J_{n-1}+{1 \over 2})$.  It is convenient to replace $(\vec{\bf
S}^{(n)})^2$ by ${\bf O}_n$ in our set of commuting operators, which can
clearly be done while still maintaining a complete set.
We can then replace the quantum number $J_n$ by $O_n$, to obtain the basis
$|\lambda,O_n,m_\alpha\rangle$.  It is easy to verify that $ \{{\bf O}_n,{\bf
s}_\alpha^{(n)} \} = {\bf S}_\alpha^{(n)}$. If we examine the effect of ${\bf
s}_\alpha^{(n)}$ on the basis $|\lambda,O_n,m_\alpha \rangle$ (where we have
defined $m_\alpha$ in the orientation corresponding to ${\bf S}_\alpha^{(n)}$),
we find that
\begin{eqnarray}
(O_n^\prime+O_n)\langle \lambda, O_n,m_\alpha| {\bf s}_\alpha^{(n)}
|\lambda^\prime, O_n^\prime ,m_\alpha^\prime \rangle \nonumber \\ = m_\alpha
\delta_{\lambda, \lambda^\prime } \delta_{O_n,O_n^\prime} \delta_{m_\alpha
,m_\alpha^\prime}
\label{eq:1}
\end{eqnarray}
Thus we see that the only non-zero matrix elements occur when $O_n^\prime=O_n$
or $O_n^\prime=-O_n$.  From this it follows that the final step in the paths of
Fig.~(\ref{fig1}) can either flip sign or else must remain the same.  Using the
relation between the $O_n$ and $J_n$ bases, this results in the selection rules
$\Delta J_n = \pm 1,0$ for ${\bf s}_\alpha^{(n)}$ acting on states in the
$|\lambda, J_n, m_\alpha \rangle$ basis. Note further that if we had choosen a
basis with $m_\beta$ instead of $m_\alpha$ in Eq.~(\ref{eq:1}) ($\beta \neq
\alpha$), the same selection rules would hold, but now the $m_\alpha$
components could be mixed by ${\bf s}_\beta^{(n)}$. In \cite{Kempe:00} it was
shown that the exchange operation ${\bf E}_{ij}={1 \over 2} {\bf I}+2 \vec{\bf
s}^{(i)} \vec {\bf s}^{(j)}$ which exchanges qubits $i$ and $j$ modifies only
the degeneracy index $\lambda$ of the $|\lambda,J_n,m_\alpha\rangle$ basis.
Because ${\bf s}_\alpha^{(j)}= {\bf E}_{jn} {\bf s}_\alpha^{(n)} {\bf E}_{jn}$,
this implies that any single qubit operator ${\bf s}_\beta^{(i)}$ can therefore
give rise to mixing of both the spin projections $m_\alpha$, and of the
degeneracy indices $\lambda$.

These selection rules must be modified for the $J_n=0$ states. $O_n=-1$ and
$m_\alpha=0$ for all $J_n=0$ states and any transitions between these states
will therefore have zero matrix element, {\it i.e.}, $\langle
\lambda,J_n=0,m_\alpha| {\bf s}_\alpha^{(n)}|\lambda^\prime, J_n^\prime=0
,m_\alpha^\prime \rangle =0$. Thus the transitions $\Delta J = 0$ are forbidden
for $J_n = 0$, and ${\bf s}_\alpha^{(n)}$ {\em must} take $J_n=0$ states to
$J_n=1$ states. Furthermore, since $\langle \lambda, J_n=0,0| {\bf
s}_\alpha^{(n)} |\lambda^\prime,J_n'=0,0 \rangle =0$, the degeneracy index
$\lambda$ for $J_n=0$ states is not affected by any single qubit
operation.

To summarize, we have shown that any single qubit operation ${\bf
s}_\alpha^{(i)}$ enforces the selection rules $\Delta J_n=\pm 1,0$
with {\em the important exception of $J_n=0$ which must have $\Delta J_n=+1$}.  The
degenerate $J_n=0$ states are therefore a quantum error detecting code for
single qubit errors \cite{Bacon:99,Kempe:00}, with the special property that
they are also the ground state of a realistically implementable Hamiltonian
\cite{Kitaev}.  The system Hamiltonian ${\bf H}_0^{(n)}$ has a ground state, for n even, with the remarkable property that all single qubit errors ${\bf s}_\alpha^{(i)}$ become dissipative heating errors.

{\em Supercoherence.}---Fig.~(\ref{fig1}) shows that for an even number of
qubits the $J_n=0$ ground state of ${\bf H}_0^{(n)}$ is degenerate.  For $n=4$
physical qubits the ground state is two-fold degenerate \cite{Zanardi:97,Bacon:99}.  This
degeneracy cannot be broken by any single qubit operator, and single qubit
operations must take the $J_4=0$ states to $J_4=1$ states, as described
above.  The system Hamiltonian ${\bf H}_0^{(n)}$ has a ground
state, for $n$ even, with the remarkable property that all single
qubit errors ${\bf s}_\alpha^{(i)}$ become dissipative heating errors.
We will call this robust ground state a supercoherent qubit. If each qubit
couples to its own individual environment, we expect that the major source of
decoherence for these ground states will indeed be the processes which take the
system from $J_4=0$ to $J_4=1$.  What kind of robustness should we expect for
the supercoherent qubit? If the individual baths have a temperature $T$, then
we expect the decoherence rate on the encoded qubit to scale at low
temperatures as $\approx e^{-\beta \Delta}$, where $\beta=(kT)^{-1}$.  At low
temperatures there will thus be an exponential suppression of
decoherence.

{\em Harmonic bath example.}---As an example of the expected supercoherence we
consider a quite general model of $4$ qubits coupling to $4$ independent
harmonic baths.  The unperturbed Hamiltonian of the system and bath is ${\bf
H}_0 ^{(4)}\otimes {\bf I} + {\bf I} \otimes \sum_{i=1}^4 \sum_{k_i} \hbar
\omega_{k_i} {\bf a}_{k_i}^\dagger {\bf a}_{k_i}$ where ${\bf a}_{k_i}^\dagger$
is the creation operator for the $i$th bath mode with energy $\hbar
\omega_{k_i}$. The most general linear coupling between each system qubit and
its individual bath is $\sum_{i=1}^4 \sum_{k_i} \sum_{\alpha}
 {\bf s}_\alpha^{(i)} \otimes (g_{i,\alpha} {\bf a}_{k_i}+ g_{i,\alpha}^*{\bf a}_{k_i}^\dagger
)$.  According to the selection rules described above we can write ${\bf
s}_\alpha^{(i)}=\sum_{(m,n) \in \cal S} {\bf A}_{i,\alpha}^{(m,n)} +{\rm
h.c.}$, where ${\bf A}_{i,\alpha}^{m,n \dagger}$ takes states $J_n=m$ to
$J_n=n$ (and acts on $\lambda$ and $m_\alpha$ in some possibly nontrivial
manner), and $\cal S$ is the set of allowed transitions ${\cal S}=\{ (0,1),
(1,2), (1,1), (2,2) \}$. In the interaction picture, after making the
rotating-wave approximation \cite{scully}, we find
\begin{eqnarray}
{\bf V}(t)= \sum_{i,\alpha,k_i,(m,n) \in {\cal S}} g_{i,\alpha}^* e^{-{i
({\Delta \over \hbar}  f(m,n)- \omega_{k_i}) t}} {\bf A}_{i,\alpha}^{(m,n)}
{\bf a}_{k_i}^\dagger \nonumber \\ + g_{i,\alpha} e^{{i ({\Delta \over \hbar}
f(m,n)- \omega_{k_i}) t}} {\bf A}_{i,\alpha}^{(m,n) \dagger} {\bf a}_{k_i}
\end{eqnarray}
where $f(m,n)=n(n+1)-m(m+1)$.  Coupling to thermal environments of the same
temperature, under quite general circumstances (Markovian dynamics,
well behaved spectral density of field modes) we are led to a master
equation (see for example \cite{scully})
\begin{equation}
{ \partial \rho \over \partial t} = \sum_{i,\alpha,(m,n) \in {\cal S}}
\gamma_{i,\alpha}^{(m,n)} {\cal L}_{i,\alpha}^{(m,n)}
[\rho]+\gamma_{i,\alpha}^{(n,m)} {\cal L}_{i,\alpha}^{(n,m)} [\rho],
\end{equation}
with ${\cal L}_{i,\alpha}^{(m,n)}[\rho]=([{\bf A}_{i,\alpha}^{(m,n)} \rho ,
{\bf A}_{i,\alpha}^{(m,n)\dagger }] + [ {\bf A}_{i,\alpha}^{(m,n)},\rho {\bf
A}_{i,\alpha}^{(m,n)\dagger}])$.  The only operators which act on the
supercoherent qubit are ${\bf A}_{i,\alpha}^{(0,1)}$.  The relative decoherence
rates satisfy $\gamma_{i,\alpha}^{(0,1)} \propto n(T)$ where $n(T)=\left[ \exp( \beta \Delta) -1
\right]^{-1}$ is the
thermal average occupation number. Thus we see, as predicted that the supercoherent qubit decoheres
at a rate which decreases exponentially as $kT$ decreases below
$\Delta$.

Finally, we note that there are additional, two-qubit errors on the system
which can break the degeneracy of the supercoherent ground state.
Such terms will arise in higher order
perturbation theory and will result in a reduced energy gap of
$g^2 \over \Delta$.  These terms will produce decoherence
%BW 05/03
rates ${\cal O}(g^2 / \Delta^2 )$ smaller than the
${\cal O}(g^2)$
single qubit decoherence rates obtained without supercoherent encoding.
In the perturbative regime, $g \ll \Delta$, this
%BW 05/03
factor therefore represents the small but finite limit to the protection
offered by supercoherence.

{\em Universal quantum computation.}---In order to be useful for quantum
computation, the supercoherent qubits should allow for universal quantum
computation. Extensive discussion of universal quantum computation on qubits
encoded in decoherence-free subsystems has been given in
\cite{Kempe:00,Bacon:99} (see also \cite{LidarPRL,DiVincenzo:00}) where it was
shown that computation on these encoded states can be achieved by turning on
Heisenberg couplings between neighboring physical qubits.  This means that we
need to add extra Heisenberg couplings to the the supercoherent Hamiltonian
${\bf H}_0^{(4)}$.  For a single supercoherent qubit these additional
Heisenberg couplings can be used to perform any SU$(2)$ rotation, {\it i.e.},
an encoded one-qubit operation. In the present scheme one would like this
additional coupling to avoid destroying the energy gap which suppresses
decoherence. This can be achieved if the strength of the additional couplings,
$\delta$, is much less than the energy gap, {\it i.e.}, $\delta \ll \Delta$.
The trade-off between the decoherence rate and the speed of the one qubit
operations can be quantified by calculating the gate fidelity $F \propto
{\delta} e^{\beta (\Delta - \delta)}$. $F$ quantifies the number of operations
which can be done within a typical decoherence time of the system. For small
$\delta$ the gates are slower, while for larger $\delta$ the gap is smaller,
resulting in a tradeoff.  $F$ is maximized for $\delta_0=k T$.  At this maximum
$F$ is still exponentially enhanced for lower temperatures. In particular, $F
\mid_{\delta=\delta_0} \propto \beta^{-1} e^{\beta \Delta}$.

Of more concern for the present scheme is how to perform computation between
two encoded supercoherent qubits.  It can be shown that using only Heisenberg
couplings, a nontrivial two encoded qubit gate cannot be done without breaking
the degeneracy of the ${\bf H}_0^{(4)}$ Hamiltonian on the two sets of four
qubits. This can be circumvented by considering a joint Hamiltonian of the
eight qubits, ${\bf H}_0^{(8)}$.  This Hamiltonian has a ground state which is
$14$-fold degenerate, including the tensor product states of the degenerate
ground state of the ${\bf H}_0^{(4)}$ Hamiltonian.  The universality
constructions previously presented in \cite{Kempe:00,Bacon:99} can then easily
be shown to never leave the ground state of this combined system.

Having shown how to perform quantum computation on the encoded qubits, it is
also apparent that the supercoherent qubit will suffer decoherence when there
is a lack of control of the Heisenberg interactions used either in constructing
${\bf H}_0^{(n)}$ or in performing a computation.  Unless the magnitude of
fluctuations in the Heisenberg interaction is larger in comparison to the bath
temperature, the resistance of the supercoherent qubit to local decoherence is,
however, unaffected by these errors.  Supercoherence, then, represents a method
for eliminating single-qubit decoherence process when superior two-qubit
Hamiltonian control is possible\cite{bacon:thesis}.

{\em Separation of control and decoherence:}---A question which naturally
arises is how the supercoherent qubits differ from encoding into a degenerate
or nearly degenerate ground state of a single physical system.  For example,
one could encode information into the nearly degenerate hyperfine levels of an
atomic ground state.  There are essentially two differences between such a
scheme and the supercoherent qubit. The first difference lies in the fact that
the degenerate ground state of a single quantum system can interact with its
environment in such a way that the coherence of the state is lost without the
bath supplying energy to the system, i.e. non-dissipatively. However, such a
mechanism cannot affect a supercoherent qubit, because all ${\bf
s}_\alpha^{(i)}$ interactions have been shown above to supply energy from the
bath to the system.  A second difference lies in the extent of efficiency in
manipulation of the system.  If a nearly degenerate ground state is used for
quantum computation, there is a tradeoff between the speed of a single qubit
gate and the decoherence rate. The limit on the speed of a supercoherent gate
is, on the other hand, related to the temperature of the bath, with an error
rate per quantum operation that scales in an exponentially favorable fashion.
Supercoherent qubits therefore obtain a separation between {\it controlled}
manipulation and {\it uncontrolled} decoherence, by making the control
mechanisms two-body interactions which the single-qubit local decoherence
cannot affect.

{\em Implementation in quantum dot grids.}---The technological difficulties in
building a supercoherent qubit are daunting but we believe within the reach of
present experiments.  In particular, these coherence-preserving qubit states appear
perfect for solid state implementations of a quantum computer using quantum
dots\cite{dots}.   Related decoherence-free encodings on $3$-qubit states were
recently shown to permit universal computation with the Heisenberg interaction
alone in \cite{DiVincenzo:00}.  The main new requirement for the supercoherent
encoding, which allows the additional exponential suppression of decoherence
not naturally achieved in decoherence-free states, is the construction of ${\bf
H}_0^{(4)}$ and ${\bf H}_0^{(8)}$.  ${\bf H}_0^{(4)}$ can be implemented by a
two dimensional array with Heisenberg couplings between all four qubits. ${\bf
H}_0^{(8)}$ poses a more severe challenge, since the most natural geometry for
implementing this Hamiltonian is eight qubits on a cube with couplings between
all qubits.  Such structures should be possible in quantum dots by combining
lateral and vertical coupling scheme. Finally, estimates of the strength of the
Heisenberg coupling in the quantum dot implementations are expected to be on
the order of $0.1$ meV~\cite{dots}. Thus we expect that at temperatures below
$0.1$ meV $\approx 1$ K, decoherence should be supressed for such coupled dots
by encoding into the supercoherent states proposed here.

{\em Acknowledgements:} We thank Guido Burkard, David DiVincenzo,
Julia Kempe, Daniel Lidar, and Taycee Lyn for useful conversations.  This work
%BW 05/03
was supported in part by the NSA and ARDA under ARO contract/grant number
DAAG55-98-1-0371.  The work of KRB is supported by the Fannie and John
Hertz Foundation.

\end{multicols}


\begin{thebibliography}{10}

\bibitem{Preskill:97a}
An entry to the quantum error correcting literature is {J. Preskill},
Proc. Roy. Soc. London Ser. A {\bf 454}, 469 (1998).

\bibitem{FTpeople}
{D. Aharonov and M. Ben-Or}, In {\em Proceedings of the Twenty-Ninth Annual ACM
Symposium on the Theory of Computing}, p. 176 (1997);  D. Gottesman, PhD
thesis, Calif. Inst. of Tech., Pasadena, CA  (1997); A. Y. Kitaev, In {\em
Quantum Communication, Computing, and Measurement}, A. S. Holevo, O. Hirota,
and C. M. Caves editors, p. 181, Plenum Press, New York, (1997);  E. Knill, R.
Laflamme, and W. H. Zurek, Science {\bf 279}, 342 (1998).

\bibitem{scully}
{M. Scully and M. S. Zubairy}, {\em Quantum Optics}, (Cambridge University
Press, Cambridge 1997).

\bibitem{Kempe:00}
{J. Kempe, D. Bacon, D. A. Lidar, and K. B. Whaley}
Phys. Rev. A, {\bf 63}, 042307 (2001)

\bibitem{Zanardi}
Decoherence-free subsystems are a generalization of decoherence-free subspaces
(which were first presented in Ref.~\cite{Zanardi:97}.)

\bibitem{Zanardi:97}
P. Zanardi and M.  Rasetti, Mod. Phys. Lett. B {\bf 11},1085 (1997)

\bibitem{Knill:00}
{E. Knill, R. Laflamme, and L. Viola}, Phys. Rev. Lett. {\bf 84}, 2525 (2000)

\bibitem{LidarBacon}
{D.A. Lidar, I.L. Chuang, and K.B. Whaley}, Phys. Rev. Lett. {\bf 81}, 2594,
(1998); {D. Bacon, D.A. Lidar, and K.B. Whaley}, Phys. Rev. A {\bf 60}, 1944
(1999)

\bibitem{Bacon:99}
{D. Bacon, J. Kempe, D. A. Lidar, and K. B. Whaley}, Phys. Rev. Lett. {\bf 85},
1758 (2000).

\bibitem{Kitaev}
A. Y. Kitaev, ``Fault-tolerant quantum computation by anyons'', LANL preprint
quant-ph/9707021; J. P. Barnes, W. S. Warren, Phys. Rev. Lett. {\bf 85}, 856
(2000).

\bibitem{LidarPRL}
{D. A. Lidar, D. Bacon, K. B. Whaley}, Phys. Rev. Lett. {\bf 82}, 4556 (1999).

\bibitem{DiVincenzo:00}
{D. P. DiVincenzo, D. Bacon, J. Kempe, G. Burkard, and K. B. Whaley}, Nature
{\bf 408}, 339 (2000).

\bibitem{bacon:thesis}
It is possible to construct supercoherent systems where the computation degrees
of freedom are seperate from the supercoherent-inducing degrees of freedom: D.
Bacon, Ph.D. thesis, Univ. Calif. Berkeley, 2001.

\bibitem{dots}
{D. Loss and D. P. DiVincenzo}, Phys. Rev. A {\bf 57}, 120 (1998).  {G.
Burkard, D. Loss, and D. P. DiVincenzo}, Phys. Rev. B {\bf 59}, 2070 (1999).

\end{thebibliography}
\end{document}